\documentclass[12pt]{article}
\usepackage[dvips]{graphicx}

\newcommand{\be}{\begin{equation}} \newcommand{\ee}{\end{equation}}
\newcommand{\ba}{\begin{eqnarray}} \newcommand{\ea}{\end{eqnarray}}
\newcommand{\nn}{\nonumber} \newcommand{\ra}{\rightarrow}

\input epsf

\begin{document}

     \centerline{\Large\bf   QCD Form Factors and  }
\centerline{\Large\bf Hadron Helicity Non-Conservation}

\bigskip
\centerline{\bf John P. Ralston$^1$ and Pankaj Jain$^{2}$}
\centerline{\bf }

\bigskip
\begin{center}

$^1$Department of Physics \& Astronomy\\
University of Kansas\\
Lawrence, KS 66045\\

$^2$Physics Department\\
I.I.T. Kanpur, India 208016\\
\end{center}
\bigskip

\noindent {\bf Abstract:}
Recent data for the ratio $R(Q)= QF_{2}(Q^{2})  /F_{1}(Q^{2})$ shocked
the community by disobeying expectations held for 50 years.  We
examine the status of perturbative QCD predictions for helicity-flip
form factors.  Contrary to common belief, we find there is no rule of
hadron helicity conservation for form factors.  Instead the analysis yields an
inequality that the leading power of  helicity-flip processes may
equal or exceed the power of helicity conserving processes.
Numerical calculations support the rule, and extend the result to the
regime of laboratory momentum transfer $Q^{2}$. Quark orbital angular momentum, an important feature of the helicity flip processes, 
may play a role in
all form factors at large $Q^{2}$, depending on the quark wave functions.
\newpage

\section{The Nature of High Energy Reactions}
 
There is an asymmetry in high energy reactions due to the Lorentz
transformation.  The spatial coordinate parallel to the boost axis is
Lorentz contracted.  The momentum fraction $x$ of
partons inside hadrons is thereby distributed over the entire possible
range $0<x<1$.  This phenomenon is dynamical, because a boost
of interacting fields is dynamical, and the $x$ dependence of wave
functions cannot reliably be calculated in perturbative QCD. Instead,
the $x$-dependence of wave functions is extracted from experimental
data.

Meanwhile the transverse spatial coordinate $\vec b$ is Lorentz
invariant.  The transverse coordinate has a certain calculability via
perturbative QCD. There is a great deal of interest and controversy
associated with the transverse coordinate in exclusive reactions.  
The transverse spatial coordinate can be probed in reactions sensitive
to the angular momentum flow.  In some reactions, the sum of the
helicities going into a hadronic reaction is automatically conserved. 
This is the case of the proton's Dirac form factor $F_{1} $ in the
high energy limit.  When the sum of the helicities is not conserved,
angular momentum conservation requires either extra constituents, or
quark orbital angular momentum.  This is the case of $F_{2}$, the
proton's Pauli form factor.

The Jefferson Laboratory has observed $Q \, F_{2}(Q^{2})/F_{1}(Q^{2})
\sim const.$ up to the highest values of $Q^{2} \sim 5.8$ GeV$^{2}$
yet measured \cite{jones,Gayou}.  The data was initially very
surprising, and the field may have reached a pivotal point in
comparison with the quark model.  For a long time it was held sacred
that $Q^{2}F_{2}/F_{1} \sim const$ at large $Q^{2}.$ This rule appears
to predate QCD. It has ancient origins in renormalization questions
involving protons as elementary fundamental fields \cite{salam}. 

Meanwhile a perturbative QCD model assuming non-zero quark orbital angular
momentum (OAM) predicted $Q \, F_{2}(Q^{2})/F_{1}(Q^{2}) \sim const$ 
\cite{Buniy,Bologna}.
A relativistic quark model prediction \cite{miller} fits the flatness
of $Q \, F_{2}(Q^{2})/F_{1}(Q^{2}) $ equally well.  These papers
countered the ancient wisdom, because they shared the common
feature of quark orbital angular momentum in the wave functions.

The hypothesis of zero OAM sometimes appeals to the non-relativistic
quark model in the rest frame.  Yet very little of high energy physics
and pQCD starts in the rest frame.  The perturbative quark wave
functions are unrestricted in angular momentum content, except for
Lorentz symmetry.  Observation of non-zero quark OAM is a leading
candidate to resolve the proton ``spin crisis'', which is the fact
that the sum of the perturbative quark spins does not equal the spin
of the proton.  Consequently the Jefferson lab's data has even broader
implications than the mystery of large $Q^{2}$ form factors.

Here we address whether $Q \, F_{2}(Q^{2})/F_{1}(Q^{2}) \sim const.$
is a transient feature of comparatively low energy experiments, or a
fact destined to persist at higher $Q^2$.  If the flatness of this
ratio is due to quark OAM, will the ratio stay flat with increasing
$Q^{2}$?  The question leads us to re-examine the roots of the 
``hadron helicity conservation'' rule \cite{BL}.  We find that pQCD itself is
rather neutral, and only gives an {\it inequality} between the powers
governing helicity-flip processes and helicity conserving ones.  As a
result $QF_{2}/F_{1} \sim const.  $ may extend to arbitrarily large
values of $Q^{2} $, without violating anything sacred.
 
\section{Definitions and Discussion}
 
\paragraph{Orbital angular momentum:}  By quark orbital angular momentum (OAM) we refer to an expansion in
$SO(2)$ representations (commonly known as $L_{z}$ states) with
quantization axis aligned along the particle 3-momentum $\vec P$.  The
eigenstates of $L_{z}$ are invariant under boosts along the $z$-axis. 
We do not use spherical harmonics, and we treat the longitudinal
coordinates as scaling variables.  Let the transverse spatial
coordinate be $\vec b=b \, (cos\phi, \, sin\phi)$.  We expand
operators or wave functions as $$\psi
=\sum_{m}e^{im \phi}\psi_{m}(x, \,b)$$
where $x$ is the Feynman light-cone fraction of $z$ (or ``+'')
momentum.\footnote{Of course perturbative wave functions are theory
constructs, and fields at finite separation are technically not gauge
invariant, in general. Observables are nevertheless gauge invariant when all the
legs of diagrams are contracted properly.} OAM can also be
re-expressed with generalized parton distributions in a manifestly
gauge-invariant formalism \cite{BNL93,Ji:1995cu}.

\medskip 

\paragraph{Chirality versus Helicity:} The proton's Pauli form factor
$F_{2}$ contains information on the orbital angular momentum of the
quarks, but it is indirect.  Strictly speaking, the amplitude $i\bar u(p', \,
s')F_{2}(Q^{2}) \sigma^{\mu\nu}q_{\nu} u(p,s)$ represents the
amplitude for {\it chirality} of the proton to flip under momentum
transfer $Q$.  The chirality (eigenvalue of $\gamma_{5}$) of light
quarks flips very little in pQCD reactions, by about $m_{q}/Q<<1$, for quarks of mass
$m_{q}\sim few \, MeV.$ The chirality of Fermions is proportional to
their helicity, with corrections of order $m /Q$.  Adding another 
constituent to the scattering is also down by $O(1/Q^{2})$. Then by this 
reasoning, at large $Q^{2} >
GeV^{2}$, it is not possible to flip either the chirality or helicity of the
proton with a virtual photon, unless there is internal quark orbital angular
momentum present to satisfy the selection rules.

\medskip 

\paragraph{What is Quark Counting?} We separate the quark-counting
model of Brodsky, Farrar, and independently Matveev et al \cite{BFMatveev}
from the
asymptotic short distance ($asd$) model of Brodsky and Lepage
\cite{BL80,BL89}. The
earlier theory is one of counting propagators and the number of
scattered constituents.  The latter theory is much more detailed, 
imposing a certain factorization of the hard reaction into components
made from the $s$-wave, $m=0$ Bethe-Salpeter wave functions.  This
does not come by listing all diagrams, and is not a feature of the
starting theory.  Instead the framework is developed by assuming the
framework and classifying terms within the assumptions.
The $asd$ approach is characterized by
taking the zero-distance limit in the first step, and replacing the rest
of the calculation by integrals over Feynman $x$ fractions using
``distribution amplitudes", or similar quantities with no transverse
information. However the Feynman rules instruct us to leave the
longitudinal and transverse integrals coupled. Perform the integrals,
and afterwards take the limit of large $Q$ (if wanted). If the two
limits are not the same (and they are not in general), then the $asd$
assumptions can fail.

The
transverse coordinates are gone in the $asd$ approach, being evaluated
within $1/Q$ of zero.  HHC follows instantly as a test of the
framework, independent of the wave functions used.  By omitting OAM
$m\neq 0$, the prediction for $F_{2} = 0$, and one cannot recover any
prediction for finite $F_{2}$ in the formalism.  As far as we know,
all $asd$ predictions of $F_{2}$ are indirect and deduced by
elimination: since $F_{1}\sim 1/Q^{4}$, and $Q F_{2}$ could not be
calculated, then $F_{2 }\sim F_{1}/Q^{2} $ must lie in the detritus
not calculated. A recent calculation \cite{BJY}, 
however, finds that within the
framework of a generalized $asd$ model $F_2\sim 1/Q^6$.
It is also possible to modify the model by
including quark mass effects \cite{crji}.  Our goal here is to
understand the limit of arbitrarily light quarks and neglecting
effects of order $m_{q}/Q$.

We separated quark counting from $asd$ because the two are not the
same theory.  Should one believe either?  There are good indications
from the scaling laws that quark-counting has some truth.  We have no
religious commitment here, and the possibility that a fraction of the
amplitude is due to ``soft physics'' must be given credit \cite{ILLS,Rady}.  But
amplitudes cannot all be soft, because the form factors are not seen
to fall exponentially with $Q$.  Indeed the quark-counting scaling
laws generally work well.  Meanwhile the inapplicability of HHC to 
$F_{2}$ or any other helicity flip reactions \cite{HelFlip} shows
we cannot use $asd$.

\medskip

\paragraph{Regarding Generalized Parton Distributions:} Generalized
parton distributions ($GPD$) appear to be an ideal way to proceed.
Lorentz covariance can be used to set up matrix elements and expressions
for the form factor in an apparently model-independent way. As far as we
know, form factors were the first instance of $GPD$, used in the paper
of Soper \cite{soper77} in 1977, which also contains a transformation to
a particular transverse spatial coordinate. We will have occasion to
revisit the conclusions of that paper in Section 3.7.

Despite our $GPD$-based predictions \cite{BNL93,Buniy,Bologna} for ratios
and the welcome rediscovery of $GPD$ in the field, we chose not to make
them the vehicle for this analysis. The reason is that $GPD$ are so
general they do not immediately contain the information there are three
quarks in the proton. To incorporate the information one can start with
wave functions and integrate out all but two quark legs. Since our
concern is precisely these integrations, we would have nothing to gain.
We caution the reader, in any event, that the method of approximating
integrals by $asd$ methods will have similar limit-interchange problems
when clothed in $GPD$ language. Diehl {\it et al.} \cite{KrollGPD}
discuss general relations, and their formulas codify a relation of $F_2$
to quark OAM.

\medskip

\paragraph{Some Familiar, But Inexact Assertions:} Suppose we have non-zero
OAM in the wave functions and we try to use the assertion of
the $asd$ approach, \ba \Delta b \Delta Q_{T}>1\ .  \label{asdregion}
\ea Here $\Delta b$ is the resolved transverse quark separation, in a
frame where the momentum transfer $Q \sim \Delta Q_{T}$ is transverse. 
We have written the relation like the uncertainty principle, to give
it a chance to be seductive.  We observe next that, merely from
continuity, a wave function $\psi_{m}$ carrying $m$ units of angular
momentum scales like $b^{m}$ as $b \ra 0$.  Then under these
assumptions each unit of orbital angular momentum of the quarks will
lead to amplitudes suppressed by a corresponding power of $1/Q$ at
large $Q$.  This familar assertion has been repeated endlessly in 
the literature, yet we will show that this type of counting does not represent
QCD. It is seductive but it is not right.

\medskip

\paragraph{Our Approach:} To address $F_{2}$ properly, one must go beyond
short-distance to restore the transverse coordinate.  This is called
``impact-parameter factorization'' \cite{impact,PR90}.  This well-justified method has
dominated recent attention in perturbative QCD 
\cite{LiSterman,Li,Bolz,Gousset,ourExclusives}. 
In impact-parameter
factorization the form factor is written as \ba F(Q^{2}) =\int ({\Pi}
dx_{i}d^{2}k_{T, \, i})\ ({\Pi}dx_{i}^\prime d^{2}k_{T, \, i}^\prime)\ 
\bar \psi(x, k_{T})H(x_{i},x_i^\prime,k_{T, \, i},
k_{T,i}^\prime;\ Q)
\psi(x^\prime, k_{T}^\prime).  \label{form1} \ea 
The impulse approximation has been
used to set the light-cone ``time'' to zero.  There are no a-priori
assumptions in Eq.  \ref{form1} about short-distance.  If Sudakov
effects are used consistently, then wave functions in Eq.  \ref{form1}
concentrate the dominant region into one which is perturbatively
calculable, {\it without assuming zero-distance as a starting point.}
We do not go to the further extreme of Ref. \cite{LiSterman} towards asserting
that the ultimate output of Sudakov effects is the $asd$ model.  We
find this unjustified.   To get that result one needs assumptions about the
wave functions that are simply not known.

\section{Calculations}

Here we present pQCD calculations  and also 
illustrate certain features that are even more general. 

\subsection{Role of The Transverse Coordinate }

 Return to the general expression Eq. \ref{form1}.  To leading order
and neglecting transverse momentum in Fermion numerators, the hard
scattering depends on differences $\vec k_{T}-\vec k_{T}'$ before and
after the hard collision.  The importance of this variable seems
rather general, because the {\it sum} of the transverse momenta are
conjugate to the overall spatial location of the hard scattering,
which by translational invariance drops out.\footnote{From
translational invariance, the hard kernel must depend on the
difference of space coordinates.  The transverse separation $b$ is
only one such difference, and the one of interest here.  Other differences
such as the longitudinal ones also occur.} In a process with a single
hard exchange (a pion or meson form factor) the transverse integrals
take the form of convolution:\ba F(Q)=\int
d^{2}k_{T}f(k_{T})\bar\psi(\vec k_{T}) ; \nn \\
f(k_{T})=\int d^{2}k_{T}'H(k_{T}-\vec k_{T}'; \,Q)\psi(\vec
k_{T}'),\nn\\ =\int {d^{2}b\over (2\pi)^2} 
e^{i\vec k_{T}\cdot \vec b} \tilde H(b;
\,Q)\tilde \psi(b, \, x).  \ea The $x$ variables were suppressed.
Consequently the expressions are diagonal in $b$, with \ba F(Q)=\int
{d^{2}b\over (2\pi)^2} dx dx' {\tilde \psi^*(b, \, x)}\tilde H(b; x,\, x',\,
\,Q)\tilde \psi(b, \, x') ,\nn \\ =\sum_{mm'm''} \int \, 
{d^2 b\over (2\pi)^2} dx
dx'd\phi
\,e^{-i(m-m'-m'')\phi} {\tilde \psi^*_{m}(b, \, x)}\tilde H_{m^\prime}(b;
x,\, x',\, \,Q)\tilde \psi_{m^{\prime\prime}}(b, \, x') \label{formb}.\ea All
contributions to OAM are explicit at this stage.  Expansion of the
hard scattering $H_{m}$ was introduced because the kernel can also
carry angular momentum and be anisotropic, via the direction of the
hard momentum $\vec Q$.  The selection rules conserving angular
momentum will come from the $\phi$ integrals.

\subsection{Gluon Exchange Kernel}

Consider the simplest one-gluon approximation to the kernel, $H =4 \pi
C_{F}\alpha_{s} /q^{2} ; \:\: q^{\mu}= xP^{\mu} -x'P^{\mu'}+\vec
k_{T}^{\mu}-\vec k_{T}^{\mu'}$ ( Fig 1).  This is written out as \ba
H(\vec k-\vec k_{T}'; x,\, x',\, \,Q) & = & 4 \pi C_{F}\alpha_{s}
\frac{1}{xx'Q^{2}+ (\vec k_{T}-\vec k_{T}')^{2}}; \nn \\ \tilde H(b;
x,\, x',\, \,Q) & = & 8 \pi^2 C_{F}\alpha_{s} 
K_{0}(\sqrt{xx'Q^{2}b^{2}}).  \label{kernel} \ea
When integrated with non-singular functions of $b$ and $x$, the
dominant region is not determined by Eq. \ref{asdregion}, but instead
the Bessel function restricts to \ba \sqrt{xx'}Q b <1.
\label{trueregion}\ea Clearly Eq. \ref{trueregion} is more accurate
than Eq. \ref{asdregion}, because the partons entering the reaction
carry $xP$, rather than $P$ and can only be scattered through
$xx'Q^{2}$ momentum transfer-squared.  This has been recognized for a
long time, and common wisdom ascribes an average value of $x \sim 1/3$
for valence quarks.  It is commonly accepted that consistent
appearance of $xx'Q^{2}$ postpones any onset of short-distance
dominance to higher values of $Q^{2}$ compared to the naive
implications of Eq.  \ref{asdregion}.

\medskip

\paragraph{A Physics Question:} We pause to question how the
``uncertainty principle'' of Eq.\ref{asdregion}, $b\sim 1/Q$, could
have misled the field.  First, it was not the uncertainty principle, but a
guess at dominant regions.  Somehow $x$, from the longitudinal
coordinate, got into a relation between transverse things.   We seek a 
physical explanation.

Suppose we stand near a fast moving charge with energy $E$, mass $m$,
at impact parameter $b$.  Relativity predicts a pulse of fields with a
time scale $\Delta t \sim b/\gamma$.\footnote{ This time scale is also
a Lorentz-contracted pancake longitudinal distance scale.  Our use of
``time-scale'' is consistent with the impulse approximation.} Since
$\gamma = E/m$ the fields depend on $\Delta t \gamma/b$.  The partons
are spread over $p_{z}=x E\sim 1/\Delta t$.  Therefore the fields depend on $
(mxb)$. This explains how the longitudinal fraction inserted itself.  In a
perturbative transition amplitude between $x, \, x'$ values,
dependence on $\sqrt{xx'} Q b$ is generic, and nearly kinematic.  The
time-scale dependence on impact parameter has been identified as
important.  The time-scale ($x$) dependence is of course set by the 
{\it non-perturbative} part of the problem, the $x$ dependence of wave functions. 

\subsection{Time Scale Smearing: Interplay of $x$ and $b$}

Exploring the integrals, the $b^{|m|} \sim 1/Q^{|m|}$ counting will
still occur, from dimensional analysis, after doing the $b$
integrations in each $x, \, x'$ integration bin.  It looks like Eq.
\ref{asdregion} and its conclusions will win after all.  What is quite
surprising, and highlighted here, is that there is no simple rule {\it
after} the $x, \, x'$ integrations are done.  The scaling indicated by
dimensional analysis is ``erased'' by the $x$-integrals under broad
conditions.  The physical origin, of course, is that the {\it
dynamical time scale distribution in the wave functions} ($x$
distributions) is not something we are priviledged to predict.  The
$x$-distribution is set by the proton itself in the quiet of vacuum
over infinite time.  The ``time-scale smearing'' destroys naive use of the
uncertainty principle, and its counting of powers.  However it does 
not destroy the use of pQCD, which always contains integration over $x$.

\subsection{The Dominant Power of $x$}

It suffices to use the kernel Eq.  \ref{kernel} to show the effects. 
Since our focus is the orbital angular momentum, we consider the
integrals with factors of $b^{|m|}$ explicit.  The selection rules from
the angular integrals $ e^{-i(m-m'-m'')\phi}$ are obvious.  We use a
Gaussian function $\tilde \Phi(b, \, x, \, \zeta)=b^{A}e^{-b^{2}/(2
a^{2})}\tilde\phi(x, \zeta)$ to represent the wave functions cutting
off large $b$, as well as a Sudakov model, described below.  The
factor of $b^{A}$ is the phase-space to find $A$ quarks close together
from naive quark-counting.  We change variables for the longitudinal
fractions to \ba x=\sqrt{xx'}, \nn \\
\zeta=x/x' \nn .\ea   We parameterize
$$\tilde\phi_{m}(x, \zeta) \sim x^{r+1}(1-x)^{r+1}\phi(\zeta),$$ as $x
\ra 1$: the $\zeta$ dependence can be left unspecified.  For later use 
parameter $r$ is called the ``dominant power of $x$''. 
In discussing the region $x\rightarrow 1$, we can expand 
$\tilde\phi_{m}(x, \zeta) $ in 
a power series. Knowing the dominant power is all that is needed for 
the arguments to go through.

\subsection{Mellin Method}

We study the large-$Q$ asymptotics by calculating a Mellin transform
$F(N)$ conjugate to $Q^{N}$: \ba F(N) &=& \int_{0}^{\infty} {dQ\over Q}
\,Q^{N}F(Q),\nn \\
   &=& \int_{0}^{\infty} db b^{|m|+A+1}\int_{0}^{1}dx \int d \zeta
   \int_{0}^{\infty} {dQ\over Q} Q^{N} \, K_{0}(xbQ)\tilde \Phi(b, \,
   x, \, \zeta).  \ea The $Q$ integral is carried out easily, \ba \int
   {dQ\over Q} \, Q^{N}K_{0}(xbQ)= 2^{-2 + N} \, (xb)^{- N} \,
   \left[\Gamma \left({\frac{ N}{2}}\right)\right]^2.
   \label{integralstep}\ea Note that $b^{-N}$ emerges just as
   expected from dimensional analysis.  If we stopped here, then
   $b^{m}$ would be suppressed by $Q^{-m}$.  
   
The other integrals are then done as \ba  & &\int_{0}^{\infty}  db  
   \int_{0}^{1} dx \, (b x)^{-N} b^{|m|+A+1} x^{r+1}(1-x)^{r+1} e^{-
   b^2/(2a^{2})}=\nn \\ & & {\frac{{2^{{\frac{N+m +A - 4}{2}}}}\, {a^{m+A
   - N+2}}\, \left[\Gamma(N/2)\right]^2\, \Gamma ({\frac{2 + m +A - N}
   {2}})\, \Gamma(2 + r)\, \Gamma(2 - N + r)}{\Gamma(4 - N + 2\,r)}}.
\label{mellinresult}
\ea
We invert the Mellin transform with a contour integral: \ba F(q) 
=\frac{1}{2 \pi i}  \int_{-i \infty} ^{i \infty} 
dN \, F(N) \, Q^{-N} . \ea  The contour
in the complex $N$ plane runs in the strip of real-$N$ where $F(N)$
converges.  Deforming the contour about singularities one-by-one
generates an asymptotic series in $Q$.   We use notation $F(Q) \leq
Q^{-P}$ to indicate dependence falling {\it at least as fast as} $Q^{-P}$. 

\subsection{HHC as Inequality}
Dependence at {\it large $Q$} now comes from singularities to the {\it
right} of the convergence strip.  There are two distinct types.

The singularities of $\Gamma ({\frac{2 + m +A - N} {2}})$ are simple
poles at $$N=2+m+A +2K , \,\:\: K=0,\, 1, \, 2\ldots $$ These are
exactly the singularities creating the naive power counting of $HHC$. 
The existence of these singularities implies \ba F(Q) \leq Q^{-m-A-2}
\:\:\:(hhc \, region \, only), \ea {\it from these singularities
alone}, and barring a zero factor canceling the poles.  As expected
$b^{m} \ra Q^{-m}$: here the $HHC$ results are reproduced.  

The dominant power $r$ locates the other singularities, seen in the
pole from $\Gamma(2-N+r)$.  These poles are independent of the $b^{m}$
dependence.  This implies an additional contribution to
power-behavior, not regulated by $bQ<1$.  The poles and power behavior
are \ba N= 2+r + K \,\:\: K=0,\, 1, \, 2\ldots \nn \\
F(Q) \leq Q^{-2-r} , \:\:\:(dominant \, power \, region)\ea {\it
regardless of the value of $m$}.  If such poles dominate, there is no
power suppression of OAM $m\neq 0$.

As a result we have the following
{\it inequality} for the large $Q$ dependence of the form factor, \ba
F(Q) \leq \frac{1}{1/Q^{c+2}} \nn \\ c=Min( \, r , \, m+A) . 
\label{ineq} \ea There is no power suppression of OAM $m\neq 0$ in
general.

\paragraph{Rule of Thumb:}  We have extracted an improved $HHC$-rule of thumb: {\it
the asymptotic $Q^{2}$ dependence is determined by the minimum of
powers $m, \, r$ in the integrands.} The rule has been deliberately
simplified to powers, suppressing logarithmic factors readily
calculated, by further evaluating the Mellin inversion.  
These logarithms are obtained from the Mellin
series and are a separate phenomenon from the asymptotic-freedom
logarithms of pQCD.   The origin of the inequality Eq. \ref{ineq}  is time-scale
smearing: due to it, the suppression of OAM expected by naive counting is
transferred to smearing inside the longitudinal wave functions, and
helicity-flips can readily compete with helicity non-flip.

\paragraph{Effects of the Large $b$ Regions} 

The singularities to the {\it left} of the convergence strip determine
a series for {\it small} $Q \ra 0$.  These singularities are traced to
the Gaussian model, Sudakov $b$-cutoff, or the large $b$ cutoff.  We are not concerned
with these singularities, showing that the large $b$ cutoff drops out
of the power-laws stated above.  However the numerical normalization
of integrals depends on the cutoff method.  It is a separate
question needing separate analysis to find the numerical fit of models
to data.   Our numerical work is presented momentarily. 

On this reasoning there is always a part of the calculation which is
``strictly perturbative''.  We may calculate the large $b$ regions
only with limited reliability: fine!  The small $b$ regions known to
be calculable behave as we claim, and are the only regions under
conceptual dispute.

\medskip

\subsection{Discussion}

As far as we know this is the first time that $HHC$ has been
observed to fail for large $Q^{2} $ form factors.  Yet previous work has
made related observations that the powers of $x$ can determine the
$Q^{2}$ dependence.  Feynman's mechanism \cite{Feynman} concentrated
entirely on the region $x \ra 1$ and ignored $b$.  Feynman's work
predated pQCD, and there are certainly regions in the $b$ integrals of
that mechanism that would not be calculable, so the model is moot.
Soper's 1977 paper \cite{soper77} has a clear statement that the {\it minimum} power
of $x$ or $k_{T}$ rules the results.  The context there was the
Drell-Yan-West \cite{DYW} relation, which connects the $x\ra 1$ behavior of
inelastic structure functions with the $Q^{2}$ dependence of the form
factor.  Soper shows that one cannot prove the Drell-Yan-West relation
deductively, but one can get an inequality and force the relation by
choosing wave functions.  

The literature is clouded here by treatments {\it assuming} that the
two regions of $x\ra 1$ and $ b\ra 0$ must give the same scaling, and
forcing a result by circular logic.  As we have just shown, the
regions are different and no general rule can be made.  Underlying
this is the fact of perturbative calculability of short-transverse
distance not being on the same footing as perturbative {\it models} of
the $x \ra 1$ dependence.  For instance Brodsky and Lepage \cite{BL80,BL89}
discuss the endpoint $x \ra 1$ contribution of the proton form factor
$F_{1}$.  The authors argue that for the limit $x_1\rightarrow 1$, pQCD
implies an $m=0$ wave function of order $\alpha_{s}^{2}$
perturbatively calculated to go like $ (1-x_{1})^{1} $.  On this basis
they find that a contribution to $F_1$ scaling like $1/Q^{4}$ {\it
independent of the powers of $b$ inside the integrals.} The $Q$
dependence of this contribution is same as that obtained from the
$b\rightarrow 1/Q$ region.  Although perturbative analysis is not a
valid approach to calculating wave functions, the result is 
consistent with ours. 

Our results are due to time-scale smearing and should not be
attributed to ``end-point singularities'' as the term is commonly
used.  Inside the hard scattering kernels in pQCD are combinations of
inverse powers of $x$ or $1-x$.  {\it We are not exploiting these
inverse powers} in the evaluation of the Mellin moments. 
Our basic point is that the $x\rightarrow 1$ limit of the 
$m\ne 0$ wave function is not known. This uncertainty leads to the
inequality given in Eq. \ref{ineq}. 
The end point singularities, when present, further enhance the end point 
contributions and expand the range of allowed values of $r$,
the dominant power of $x$, 
for which there is no power suppression of OAM. 
We are also not
concerned with endpoint wave functions (a la CZ) for the same reason:
their relevance or lack of it is another subject.   

To summarize the logic so far, for broad classes of $x$ dependence 
of the wave functions, which is unknown,
the non-rule of HHC is transformed to a rule of HHNC, hadron helicity
non-conservation.  This is a new asymptotic prediction, and proves
that $Q \, F_{2}(Q^{2}) /F_{1}(Q^{2})\sim const$ can be an outcome of
the theory up to the highest $Q$.  To be equally fair, different broad
classes of wave functions give $Q^{2 } \, F_{2}(Q^{2})
/F_{1}(Q^{2})\sim const$.  Our primary accomplishment enlarges the
sphere of allowed possibilities.  Measurements are still needed
to determine what protons are.

\section{Numerical Studies, Sub-Asymptotic}

The scaling rule $Q \, F_{2}(Q^{2}) /F_{1}(Q^{2})\sim const$ describes
experiments at laboratory $Q$-values very far from asymptotic.  We
explored this region numerically.  

\subsection{Studies with the Pion}
 
The pion provides a simple test
system.  We test the dominant integration regions by inserting factors of $b^{m} $ in
the integrands, corresponding to $m$-units of OAM. The moment $<b(Q)>_\pi$ is
defined by
\begin{equation}
<b(Q)>_\pi = {\int dxdx^\prime d^2 b b {\cal F}_\pi(Q, \, x, \,  x^\prime, \, b)
\over F_\pi(Q^2)}.
\end{equation}
The pion form factor 
$F_\pi(Q^2)$ is given by
\begin{equation}
F_{\pi} (Q^2) = \int dxdx^\prime d^2 b  {\cal F}_{\pi} (Q, \, x,\, x^\prime,\,  b)
\end{equation}
with the kernel ${\cal F}_\pi(Q, \, x,\, x^\prime,\, b)$  defined by
\begin{equation}
{\cal F}_{\pi} (Q, \,  x,\,  x^\prime,\,  b) = \int dx dx^\prime db b\phi(x^\prime)
\alpha_s(\mu)e^{-S(x,\,  x^\prime,\, b,\,  Q)}K_0(\sqrt{xx^\prime}bQ) \phi(x)\Phi(b)\ .
\label{calFpi}\end{equation}
Here $S(x,\, x^\prime,\,  b,\,  Q)$ is the Sudakov form factor.  A model soft
wavefunction $\Phi(b)=\exp(-b^2/2a^2)$ is also included, where the
parameter $a=1/\Lambda_{QCD}$.  

\begin{figure}
\includegraphics[scale=0.8]{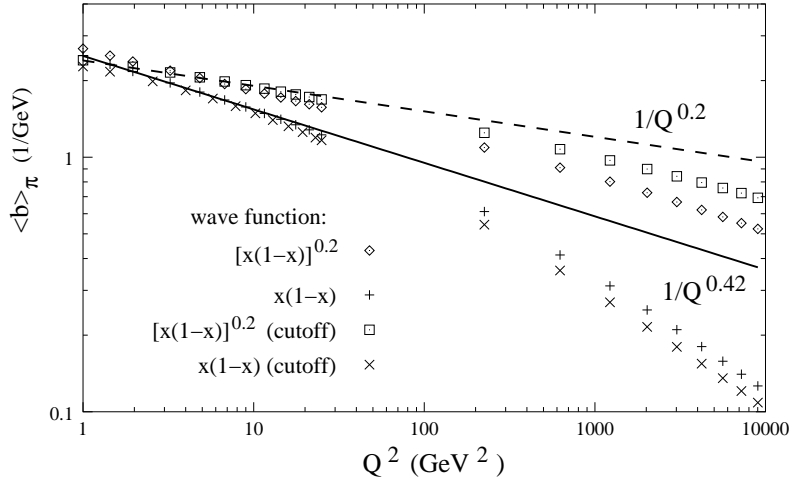}
 \caption{The moment $<b(q)>_\pi$ of the pion form factor kernel, as
 defined in the text, using the $[x(1-x)]^{\delta}$ form of wave
 function for $\delta =0.2, \, 1$. The moment decreases with $Q$ much
 slower than the $1/Q$ behaviour expected in the $asd$ HHC model. 
 Results are shown with and without including the Sudakov effects.
 The solid $(1/Q^{0.42})$ and the dashed $(1/Q^{0.2})$ line 
 represent a simple power law fit at small $Q^2$ for $\delta=1$ and
 $\delta=0.2$ respectively.
 }
\label{PionbMoments}
\end{figure}

In Fig. \ref{PionbMoments} we plot the moment $<b(Q)>_{\pi} $ using a wave
function of the form $$\phi(x) \sim [x(1-x)]^\delta,$$  We study values
of $\delta=0.2,1$, with and without the Sudakov form factor.  In
the latter case the momentum scale $\mu =Q/4$ of the strong coupling
is imposed, and the $b$ integrals are cut off at
$b=1/\Lambda_{QCD}$.  From the Figure it is clear that for $Q^2<100$
GeV$^2$ the moment falls much slower than the $asd$ prediction of
$1/Q$, regardless of the $x$-dependence of the wave function used.

\subsection{Studies with the Proton}

\begin{figure}
\includegraphics[scale=0.8]{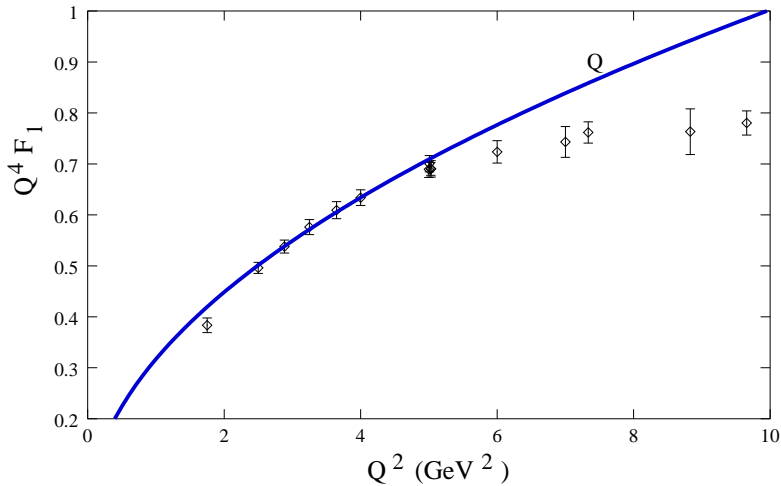}
\caption{The proton form factor $F_1$ for moderate $Q^{2}$,  extracted using the Jlab data
for $G_E/G_M$ and the SLAC data for $G_M$.  The ratio $G_E/G_M$ is
obtained from the parameterization \cite{Gayou} $\mu_p {G_E\over G_M} =
1 - 0.13(Q^2-0.04)$. The solid line represents $F_1\sim 1/Q^3$.}
\label{fig:data_F1}
\end{figure}

We turn to the proton form factors.  We note that the JLAB data for
$QF_2/F_1$ is flat even below the $Q$-range where $F_1 \sim 1/Q^4$
begins to fit.  In Fig.  \ref{fig:data_F1} we show a plot of 
$Q^4 F_1$. The solid line in the plot corresponds to the behaviour 
$F_1\sim 1/Q^3$. It is clear from the plot that the scaling $F_1\sim 1/Q^4$
is seen only for $Q^2\ge 6$ GeV$^2$, which is larger than the momentum
regime explored at JLAB so far.
This is cause for concern.  It is
nevertheless entirely possible that the scaling observed in the ratio
is not overly sensitive to $Q^{2}$, and will continue to larger $Q^{2}$. 
We investigate this in greater detail numerically.

\begin{figure}
\includegraphics[scale=0.8]{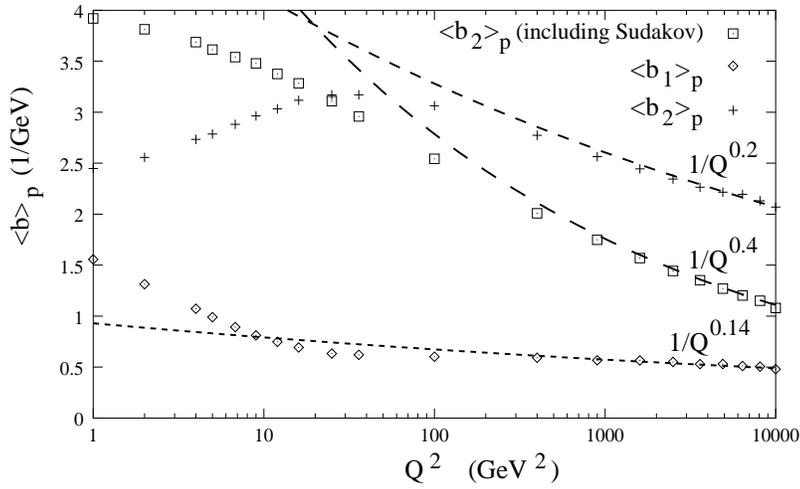}
\caption{The moment of the transverse separation $b_1$ and $b_2$
for proton using the wave function of the form $(x_1x_2x_3)^{0.5}$.
The $<b_2>_p$ moment is shown both with and without including the sudakov 
form factor.
The small dashed $(1/Q^{0.14})$, medium dashed $(1/Q^{0.2})$ and 
the large dashed $(1/Q^{0.4})$ lines
represent simple power law fits at large $Q^2$.
}
\label{fig:moments_proton}
\end{figure}

To probe the dominant integration regions, we turn to calculating $b$-
moments of the proton form factor kernel.  
These are multidimensional
integrals of which our analysis in the previous section determines but a
low-dimensional strip. We spare the reader a listing of dozens of
Feynman diagrams and substantially complicated kernels listed in the
literature \cite{Li}. The form factor $F_1$ in impact parameter
coordinates can be written symbolically as \cite{Li}
\begin{eqnarray}
F_{1}(Q^{2})&=&\sum_{j=1}^2\frac{4\pi}{27}
\int_0^1 (d x)(d x')\int_0^{\infty}
b_1 d b_1 b_2 d b_2 \int_0^{2\pi} d\theta [f_{N}(w)]^{2}
\nonumber \\
& &\times {\tilde H}_j(x_{i},x_{i}',b_i,Q,t_{j1},t_{j2})\,
\Psi_{j}(x_{i},x'_i,w)
\nonumber \\
& &\times \exp\left[-S(x_{i},x_{i}',w,Q,t_{j1},t_{j2})\right]\; ,
\label{F1kernel}
\end{eqnarray}
with $(dx) = dx_1dx_2dx_3\delta(1-x_1-x_2-x_3)$.  The variable
$\theta$ is the angle between ${\bf b}_1$ and ${\bf b}_2$ and $x_i$
and $x_i^\prime$ refer to the initial and final $x$ variables.  The
expressions for the hard scattering $\tilde H_j$, the Sudakov form
factor $S$, and function $\Psi_i$ are given in Ref. \cite{Li}.   Now 
defining
\begin{equation} 
F_{1}(Q) =\int b_1 d b_1 b_2 d b_2 \,   {\cal F}_{P}  \, 
\end{equation}
from Eq. \ref{F1kernel}, we define moments $<b_{j} (Q)>_{p}$ as follows:
\ba <b_{j} (Q)>_{p} =\frac{ \int b_1 d b_1 b_2 d b_2 \,   {\cal 
F}_{P}     b_{j}  }{ F_{1}(Q^{2})}  .
\ea

The function $\Psi_i$ is where the linear combinations of the products
of initial and final $x$ wave functions are found.  The most singular
part of the kernel in the limit $x_1\rightarrow 1$ and
$x_1^\prime\rightarrow 1$ is obtained from the $\tilde H_1\Psi_1$
term, which is of the form,
\begin{equation}
\tilde H_1\Psi_1 \sim {K_{0}\left(\sqrt{(1-x_{1})(1-x_{1}')}Q b_1\right)
K_{0}\left(\sqrt{x_{2}x_{2}'}Q b_2\right)
\phi(x_i)
\phi(x_i^\prime)\over (1-x_1)(1-x_1^\prime)}
\end{equation}
Here $K_{0}$ is the modified Bessel function of order zero. 

For the test we explore a wave function $\phi(x_{i}) $ given by
$$\phi(x_{i}) \sim (x_1 x_2 x_3)^\delta\ .$$  The numerator in $\Psi_1$
is then proportional to $(x_1 x_2 x_3 x_1^\prime x_2^\prime
x_3^\prime)^\delta$.  The Bessel functions imply that in the limit of
large $Q$, $(1-x_1)(1-x_1^\prime)b_1^2 \rightarrow 1/Q^2$ and
$x_2x_2^\prime b_2^2\rightarrow 1/Q^2$.  As long as $\delta \le 0.5$
the time-scale smearing will dominate.  For $\delta=0.5$ we get
$F_1\rightarrow 1/Q^4$ even though a dominant region in $b_1$ and
$b_2$ is independent of $Q$ !  For $\delta \le 0.5$ the moments of
$b_1$ and $b_2$ should also have a region independent of $Q$.

We check these predictions by performing the calculation using
$\delta=0.5$.  We ignore the Sudakov form factor for this test,
because it is a side issue.  We evaluate the strong coupling at $Q/4$.
In figure \ref{fig:moments_proton} we plot the moment of the transverse
separations $b_1(Q)$ and $b_2(Q)$ as a function of $Q$. 
We find that for $Q^2>100$,
the moments $<b_2>_p$ and $<b_1>_p$ fall asymptotically as
$1/Q^{0.2}$ and $1/Q^{0.14}$ respectively, 
in agreement with
our analytic expectations.  If we include the Sudakov form factor the
moment $<b_2>_p$ falls as $1/Q^{0.4}$, asymptotically.  This is only
a slightly stronger decay compared to the earlier case.  In contrast to
earlier expectations, the Sudakov form factor does not much suppress
the importance of the end-point region.

\begin{figure}
\includegraphics[scale=0.8]{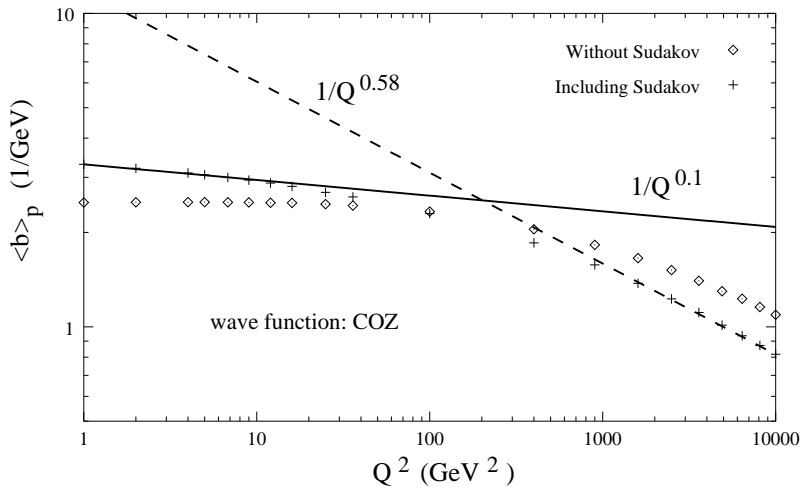}
\caption{The moment of the transverse separation $<b_2(Q)>_p$
for the proton form factor kernel using the COZ wave function \cite{COZ}.  
The solid $(1/Q^{0.1})$ and dashed $(1/Q^{0.58})$ lines
represent simple power law fits at small and large $Q^2$ respectively.
The dependence on $Q$ is much 
weaker than predicted by $asd$ relations, and supports a flat 
prediction for $QF_{2}/F_{1} $ at JLAB momentum transfers }
\label{fig:proton_b2}
\end{figure}

\begin{figure}
\includegraphics[scale=0.8]{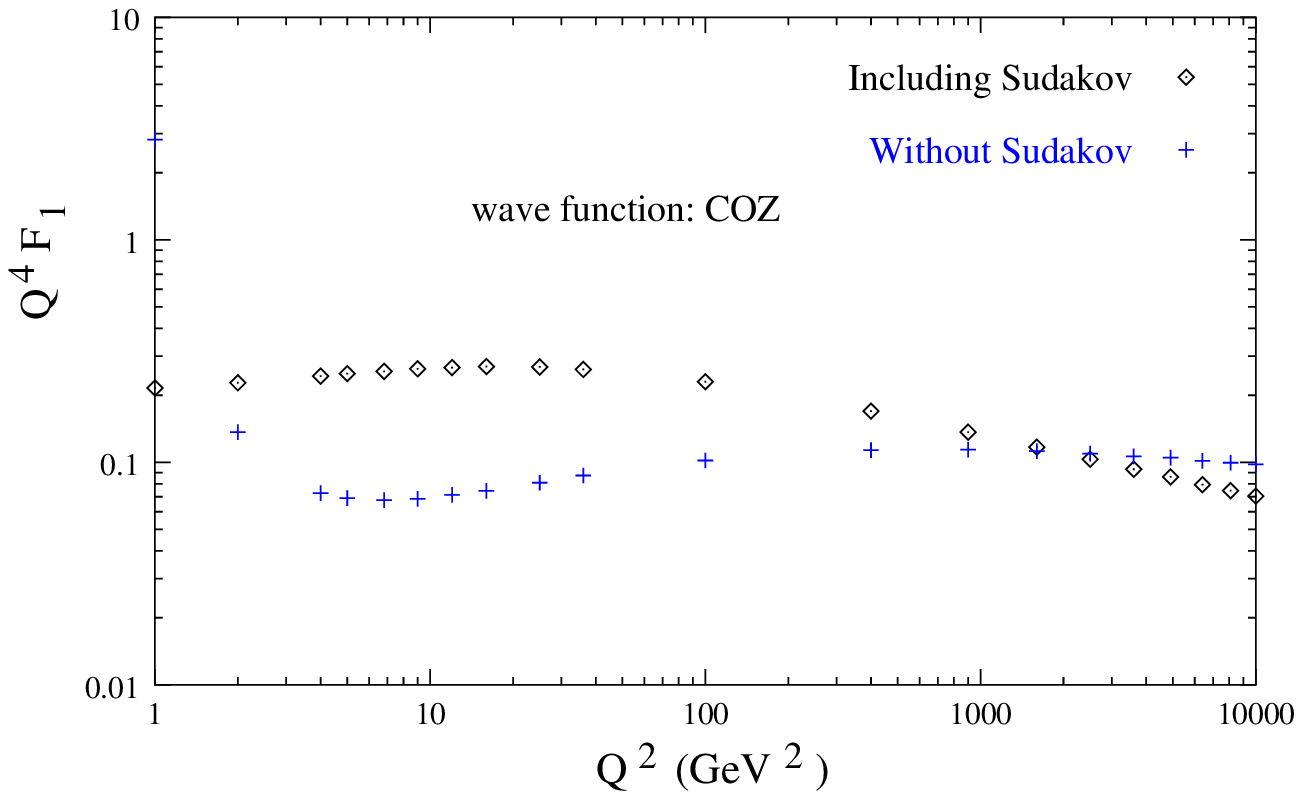}
\caption{The $Q^{2} $ dependence of the calculated proton form
factor $F_1$ using the COZ wave function \cite{COZ}.  Results are shown
both with and without including the Sudakov form factor.  Details are
in the text.}
\label{fig:proton_F1}
\end{figure}

The results with end point dominated COZ \cite{COZ,CZ}
$x$-dependence are similar. In this
case, as shown in Fig. \ref{fig:proton_b2}, we find that the $<b_2>_p$ 
decays very slowly with $Q$ for a wide range of $Q$.
This is quite interesting, because, as shown in Fig. \ref{fig:proton_F1}, 
the form factor
$F_1$ itself does not show good $1/Q^{4}$ scaling in this momentum
regime, if the sudakov form factor is not included.  
The calculation indicates that the scaling seen in the
moment, and hence the ratio $QF_2/F_1$, is more general than that seen
in $F_1$. This is not totally unexpected in a ratio.
 At very large $Q$
it starts to fall faster, but only as $1/Q^{0.6}$: well below the supposed
$1/Q$ rule of $asd$ assumptions. 

  \subsection{The Ratio $QF_{2}(Q^{2}) /F_{1} (Q^{2}) $ }
  
Based on these studies we come to a prediction for the ratio
$QF_{2}(Q^{2}) /F_{1} (Q^{2}) $.  This prediction is based on 
power-counting, and the dominant integration regions probed by the 
moments.  It is as complete a prediction as now possible, taking into 
account that one unit $\Delta m=1$ of orbital angular momentum is 
needed for $F_{2}$ to proceed for $Q^{2} >>$ GeV$^{2}$. 
Up to small and model dependent corrections, of order $Q^{0.05}$ , the
power counting
gives the scaling behavior but not the normalization of the data, via
the relation \ba <b_2(Q)>_{p} \sim \frac{Q \, F_{2}(Q^{2})}{ F_{1} (Q^{2}) } 
\sim   {\rm constant}. \ea 

What if this prediction fails?  It can fail, according to our 
asymptotic studies, if the dominant power $r$ is ``large''.  Large 
$Q^{2}$ studies of $F_{2}$ not only probe quark OAM, but also they can 
tell us details of the dominat $x$ power associated with $m \neq 0.$   

\subsection{Positivity of $G_{E}  $ }

In the Appendix of a 1959 {\it Reviews of Modern Physics}
article \cite{yennie} there are listed two form factors denoted $A$ and
$B$, now called $G_E$ and $G_M$, which are linear combinations of
Rosenbluth's: \ba G_M = F_{1}+ \kappa F_{2}; \nn
\\ G_E =F_{1}- \kappa \tau F_{2}; \nn \\ \:\:  \tau=\frac{Q^{2}}{4
m_{p}^{2}} \label{sachs}\ea
and $\kappa$ is the anomalous magnetic moment.
For some reason these are called the 
Sachs form factors \cite{Sachs}.  Yennie {\it et al} gave \cite{yennie} the
alternative linear combinations simply to emphasize that the
definitions of form factors was arbitrary. 

It is interesting to observe that if the trend of $QF_{2}(Q^{2})
/F_{1} (Q^{2}) \sim const$ continues, the value of $G_{E} $ will reach
zero, and cross to a negative value.  One wonders \cite{perdrisat}
whether there is a physical significance and a barrier to this unusual
occurence.  
The meaning of $G_{E}$ is angular momentum $\Delta
J_{z}=0$ spin-preservation in the Briet frame. If $G_{E}=0$ the
proton spin must flip in the scattering, which seems industrially
useful.\footnote{Quantum computing, which often invokes high-speed,
high-precision spin-flips, may or may not ever need the physics of
multi-GeV spin flips, but it stands ready.} We find nothing special
about this.  The amplitudes in the spacelike region cannot be limited
further than general principles of analyticity, Lorentz and gauge
invariance, and so on.  It is perfectly consistent to arrange timelike 
discontinuities so that the sign change of $G_{E}$ occurs without 
violating anything holy.  

\section{Concluding Remarks}
 
Our calculations show that the (newly revised) asymptotic behaviour of
powers of $b$ is achieved only at very large $Q^{2} $.  Physics has
many asymptotic predictions, which by construction have suppressed
information needed to know when they will apply.  To repeat: methods
inherent in asymptotic prediction strongly tend {\it not} to tell you
where the prediction will be valid.  In QCD it was early thought the
``asymptotic'' regime had to occur at energies very large compared to
$\Lambda_{QCD}$, the strong scale.  This was built wrongly into dogma. 
Very often {\it asymptotically large logarithms} are needed to justify
some approximation.  We have shown here that the scales of asymptopia
for form factors are vastly beyond experimental comparisons. 
Asymptopia is effectively meaningless, because all the new physics of
very high energies has been left out.

Meanwhile, the finite $Q^{2}$ effects of OAM are hardly suppressed at
all.  They are suppressed in explicit calculations {\it by even less
than the revised asymptotic behavior.} The pyramid of assumptions that
the $s$-wave distribution amplitudes are meaningful falls into very
grave doubt.  To put this more directly, we don't have a reason to
use a distribution amplitude any more.

The usual approach to pQCD, in which wave functions are assumed to be
unknown, predicts HHNC: {\it hadron-helicity non-conservation}.  There
is an inequality that helicity-flip processes do not lead but can have
equal power with helicity non-flip processes.

The experimental observation of $R(Q)=Q\, F_{2}/F_{1}$ appears to have a
dual meaning.  The ratio is a very robust quantity, which remains flat
even in the regime where $F_{1}$ is not clearly dominated by
quark-counting.  The same ratio is an important asymptotic quantity,
which says things about the wave functions.  The
common notion that proton wave functions are ``cubic'', namely going
like $x_{1}x_{2}x_{3}$, is ruled out for wave functions calculating
$F_{2}$.  Applications of the CZ wave functions are ruled out for
$F_{2}$, if the ratio $R$ continues to be flat at large $Q^{2}$.

Predictions for protons are mirrored in predictions of the same type
for neutrons.   If the flat $R(Q)$ ratio is followed for neutrons, then the
hallowed Galster fits \cite{Galster} to neutron data will eventually
fail \cite{perdrisat}. We predict a flat $R$ ratio for neutrons on the
basis of isospin independence of QCD.

The study of polarization transfer in large nuclei, $\vec e(A, e')\vec
p'$ should yield further information.  If the existing form factors
are dominated by $asd$, which we do not believe but is still
worth testing, then nuclear filtering will not make the distances
shorter, and the ratio $R$ should be flat with $A>>1$.  The regime of
small $A \sim 10$ has no advantages and many complications due to
few-body effects, and predictions are more difficult.  If the existing
form factors are dominated by quark mass effects, then nuclear
filtering will not make any difference and the ratio $R$ should again
be flat.  If quark OAM is responsible for the flatness of $R$ as we
believe, then filtering will kill the large transverse extent of large
OAM \cite{physRep,Dourdan}, $F_{2}$ should be depleted relative to $F_{1}$, and $R$ will
decrease with $A>>1$.  We intend to dedicate a study to quantifying
these predictions.

JLAB has made a pivotal experimental discovery which will be a
permanent subject of discussion.  The experimentally observed flatness
of $QF_{2}/F_{1}$ is a signal of substantial quark orbital angular
momentum in the proton.  Higher momentum transfer measurements would
be helpful in confirming this interpretation.  The numerical size of
$R$ cannot be converted directly to a wave function, because it is
only a single number.  But the value of $R$ can rule out models which
omit quark OAM.   Further studies at higher $Q^{2}$ may separate
constituent quark models with OAM \cite{miller}, which tend to have a scale (the
quark mass) forcing turn-over of $R$ with increasing $Q^{2}$ .

{\bf{ Acknowledgements: }} Work supported by Department of
Energy Grant Number DE-FG03-98ER41079.

\end{document}